\documentclass[10pt,conference]{IEEEtran}
\IEEEoverridecommandlockouts

\usepackage[pass]{geometry} 
\newgeometry{top=0.75in, bottom=1.04in, left=0.673in, right=0.673in}

\usepackage[cmex10]{amsmath}
\usepackage{amssymb,bm}
\usepackage{wasysym}
\usepackage[pdftex]{graphicx}
\usepackage{color}
\usepackage{bm,cite}
\usepackage{tikz}
\usepackage{mathrsfs}
\usepackage{multicol}
\usepackage{multirow}
\usepackage{soul}
\usepackage[detect-all]{siunitx}
\usepackage{booktabs}
\usepackage{multirow,tabularx}
\usepackage{enumitem}
\usepackage{algorithm}
\usepackage{algorithmicx}
\usepackage{algpseudocode}
\usepackage{subfigure}

\begin{document}

\title{Wideband Direct Satellite Uplink Enabled by  Pilot-less Sparse Superposition Codes}
\author{
 \IEEEauthorblockN{Alberto G. Perotti$^*$,  Branislav M. Popovi\'c and Renaud-Alexandre Pitaval} 
\IEEEauthorblockA{Shannon Research Center,  Huawei Technologies Sweden AB \\
Email:\{alberto.perotti@polito.it, branislav.popovic@huawei.com,  renaud.alexandre.pitaval@huawei.com\}}
\thanks{$^*$A. Perotti is now with the Department of Electronics and Telecommunications of Politecnico di Torino, Italy.} 
}

\maketitle

\begin{abstract}
Direct satellite uplink is severely constrained by limited link budgets, which hinder the exploitation of wideband resources, and ultimately limit the throughout. 
This paper presents a pilot-less coded modulation scheme based on sparse superposition coding (SSC) to enable efficient wideband usage in coverage-limited scenarios. This scheme  
leverages the structured Zadoff-Chu quasi-orthogonal (ZC-QO) dictionary to support scalable transmission. To address decoding complexity, the SSC transmitted signal embeds root index information via indicator sequences, allowing the receiver to restrict the decoding search space. 
In addition, a multi-codeword transmission framework with repetition and stop-feedback is developed, enabling reliable communication and better  resource utilization.  
 Simulation results show that the proposed scheme achieves throughput gains compared to a more conventional narrow-band multi-dimensional constellation-based approach. 
\end{abstract}

\section{Introduction}
\label{sec:Intro}
In non-terrestrial networks (NTN), direct uplink (UL) transmission 
is fundamentally constrained by a limited link budget, resulting in poor coverage and low throughput.  
This is primarily due to the combination of low user equipment (UE) transmit power and antenna gain, large propagation distances, and mobility of both the UE and the satellite. In particular, the UE transmit power -- typically around 23 dBm for commercial devices -- is insufficient to compensate for the severe free-space path loss inherent to satellite links. 
As a result, the received signal-to-noise ratio (SNR) is generally very low and may further degrade under occasional deep fading.  
Ensuring reliable transmissions at relevant throughput under such conditions is thus highly challenging. 

In multicarrier systems such as orthogonal frequency-division multiplexing (OFDM),  
increasing the data rates can in principle be achieved by allocating larger bandwidth.  
Although satellite uplink bands may offer tens of megahertz of spectrum, power-limited UEs cannot fully exploit this resource. Expanding the bandwidth reduces the transmit power per frequency resource, and thus SNR. This is especially critical for cell-edge direct satellite links, where UEs inevitably operate at saturated power. Consequently, only narrowband transmission is practical, limiting achievable throughput, whereas  UEs with more favorable channel conditions or higher-power terminals such as very small aperture terminal (VSAT) can utilize wider bandwidth more effectively, and deliver larger throughput.

Repetition is a widely used coverage enhancement technique, and an important component of 5G NTN to cope with the severe path loss  
inherent  to  satellite links.  
With multiple retransmissions of the same data block, the receiver can combine their energy to enable decoding at very low SNR.  
Due to the large round-trip time in satellite systems, hybrid automatic repeat request (HARQ) feedback may lead to stalling, so a key adaptation made in 5G NTN was to allow disabling of the HARQ feedback and let the network rely directly on many blind repetitions~\cite{QC_WiSEE23}.  
In pilot-based schemes, maintaining accurate channel estimation at very low SNR requires increasing the number of pilot signals.   
With repeated transmission, demodulation reference symbols (DMRS) can be bundled across repetitions to perform joint channel estimation~\cite{ICCT20}, improving estimation accuracy and thus decoding performance. However, for DMRS bundling to work, the UE must maintain a phase continuity across the bundled repetitions which is difficult to maintain in NTN due to Doppler shifts and timing drifts induced by satellite motion.

In~\cite{bib:ChePitDirectSatellite}, we investigated pilot-less transmission for direct satellite access in coverage-limited scenarios using pilot-less multi-dimensional constellations (MDC)~\cite{MDC_Forney1989,Hochwald2000,Pitaval2020,Qin2020,Pitaval2021}.
By avoiding explicit channel estimation, MDC eliminates the dependency on channel estimation accuracy and removes pilot overhead, allowing all time–frequency resources and transmit power to be allocated to data.  
In~\cite{bib:QinPitPUCCH,bib:ChePitDirectSatellite}, MDC was shown to bring performance gains in low-SNR scenarios of small bandwidths of a few physical resource blocks (PRBs) over a 5G-like transmission with LDPC code and QPSK modulation.  
However, extending this MDC-based approach to wider bandwidths remains challenging. Maintaining spectral efficiency requires the MDC size and dimensionality to grow exponentially with bandwidth, leading to prohibitive decoding complexity. Consequently, MDC approach does not scale efficiently to large bandwidths.

Large modulated block transmission is conventionally supported using bit-interleaved coded modulation (BICM), which combines  binary error correction coding with one-dimensional modulation. 
In BICM, channel coding and modulation are effectively decoupled, and long block codes are easily decodable with conventional binary  codes.  
As a result, BICM readily scales to wideband operation.  
However, the near-capacity performance of BICM relies on the existence of Gray labeling in the underlying modulation~\cite{bib:BICM}.  
Such labeling generally does not exist for MDCs~\cite{Qin2020}, making bit-interleaved coded  MDC inefficient~\cite{Qin2023}.

Coded modulation schemes that map information directly to sequences of symbols, without relying on bit-to-symbol labeling,  
have been developed prior to, and concurrently, to BICM. 
Among them, sparse superposition codes (SSC) were introduced in~\cite{BarronTIT12} and shown to achieve capacity in Additive White Gaussian Noise (AWGN) channel under maximum likelihood (ML) decoding. SSC have been considered for high-reliability transmission in~\cite{SVC_TWC18},   
 where sparse superpositions of MDC vectors are constructed using randomized dictionaries and decoded via matching pursuit (MP) algorithms.  
Moreover, SSC is amenable to non-coherent decoding~\cite{PilotlessSVC19},   
enabling pilot-less transmission.   
While deterministic dictionary constructions, e.g., based on Reed–Muller sequences~\cite{bib:CalHowJaf2010JSTSP}, were  briefly discussed in~\cite{SVC_TWC18}, their performance was not evaluated. 
In~\cite{bib:QOSSC}, we introduced quasi-orthogonal SSC (QO-SSC),  
where the dictionary is structured as a collection of multiple orthogonal subsets, using constructions based on Kerdock codes and Zadoff-Chu sequences.  
QO-SSC demonstrated  competitive performance for short codewords at low rates,   
outperforming the best conventional coded modulation schemes such as polar-coded QPSK. 
However, conventional sparse decoding approaches  based on full dictionary correlation as in MP-type algorithms lead again to  prohibitive complexity as the dictionary grows.

In this paper, we propose a pilot-less SSC scheme 
to enable wideband direct satellite uplink in coverage-limited scenarios. 
The proposed method overcomes  scalability limitations of MDC while retaining the benefits of non-coherent transmission.  
Specifically,   
 we introduce a  ZC-based QO-SSC (ZC-QO-SSC) scheme  which leverages the structure of the ZC-QO dictionary. 
To address the inherent decoding complexity of long SSC codewords, we propose a novel mechanism that embeds root index information within the transmitted signal using superposed indicator ZC sequences. 
This allows the receiver to restrict the search space to the indicated orthogonal subsets, significantly reducing decoding complexity. In addition, these indicator sequences are used to obtain coarse channel estimates that are exploited to improve decoding performance of the superposed data sequence.  
Furthermore, we extend the proposed scheme to a multi-codeword transmission framework with controlled repetitions. 
The coarse channel estimates are reused to enable coherent combining across repetitions, enhancing decoding performance.  
To avoid excessive repetition overhead, a stop-feedback mechanism is used to terminate transmissions once successful decoding is achieved, thereby improving throughput and avoiding HARQ stalling in NTN.  
Simulation results demonstrate that the proposed scheme enables efficient wideband operation, achieving a maximum throughput of 64 kbps, corresponding to more than twice that of the MDC-based narrowband solution in~\cite{bib:ChePitDirectSatellite}.

\section{Preliminaries on Sparse Superposition Coding}
\label{sec:QOSSC}

\subsection{SSC Encoding}

The SSC encoder maps an information message ${\bf m}$ of length $N_{\rm info}$ bits
to a SSC codeword ${\bf c}$ of $P$ complex symbols according
to the following:
\begin{equation}
	\label{eq:QOSSCenc}
	{\bf c} = \frac{1}{\sqrt{L}} {\bf F} {\bf v}
\end{equation} 
where ${\bf F}$ is a $P\times N$ sparse superposition dictionary.
${\bf v}$ is a sparse binary column vector of length $N$ and Hamming weight $L\ll N$, where the positions of the $L$ non-zero elements are determined
by the information in $\bf m$ according to a certain sparse mapping ${\cal S}$:
\begin{equation}
{\bf v} = {\cal S}({\bf m}).
\end{equation}

The sparse mapping is not crucial for the performance -- the only relevant property
is its sparsity. As a simple practical mapping, we divide vectors $\bf m$ and $\bf v$
into $L$ non-overlapping sections of equal lengths (assuming that both $\bf m$ and
$\bf v$ have lengths multiple of $L$).
The bit values in the $l^{\rm th}$ section of $\bf m$ are converted to decimal and
used as the index of the single nonzero element in the corresponding section of $\bf v$.
Therefore, the number of information bits that can be mapped to a SSC codeword is
\begin{equation} \label{eq:Ninfo}
	N_{\rm info} = L \left\lfloor\log_2 (N/L)\right\rfloor. 
\end{equation}

MDC can be seen as a special case of SSC with $L=1$ for which then the message size is $N_{\rm info}= \log_2 (N)$, corresponding to the size of the codebook.

We are interested in OFDM transmission consisting of $N_{\rm OS}$ OFDM symbols  and $N_{\rm SC}$ 
subcarriers; each group of 12 consecutive subcarriers is referred to as a PRB. The total number of time-frequency resource elements (REs)   is therefore
$ M = N_{\rm OS} N_{\rm SC}$. 
The codeword $\bf c$ may be punctured/extended by removing/repeating some
of its elements according to a predefined pattern if the number of available
channel resources and the codeword length do not match.
The punctured/extended codeword of length $M$, denoted as ${\bf c}'$,   
is mapped to the REs of the time-frequency resource grid, e.g.,  
following a frequency-first order common in 5G NR.

\begin{figure}[t]
	\centering
	\resizebox{.95\hsize}{!}{
		\includegraphics[clip=true,trim=8.cm 9.9cm 7.9cm 9.5cm]{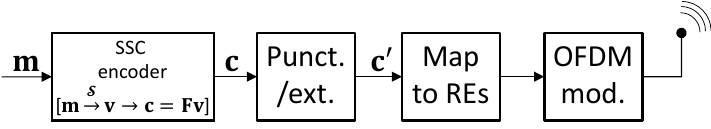}}
	\caption{SSC transmitter.}
	\vspace{-0.1cm}
	\label{fig:QOSSCenc}
\end{figure}

Figure~\ref{fig:QOSSCenc} shows a block scheme of a SSC transmitter. 

\subsection{SSC Decoder}
After demodulation, the received signal can be written as
\begin{equation}
	{\bf y} = {\bf h} \odot {\bf c} + {\bf n}
\end{equation}
where $\bf h$ is a vector of fading coefficients, $\odot$ is the Hadamard product, and  $\bf n$ is an AWGN vector.

Common SSC decoders based on variants of MP algorithms as in~\cite{SVC_TWC18,bib:QOSSC} relies on correlations of the received vector with the SSC dictionary, typically through multiple iterations.   
Assuming a single iteration for simplicity of the discussion, the received codeword $\bf r$ is decoded by computing
the correlations with all the columns of $\bf F$ as
\begin{equation}
	\label{eq:projection}
	\bm{\xi } = {\bf F}^{\rm H} {\bf y}. 
\end{equation} 
Based on the correlation vector $\bm{\xi }$, 
an estimated $L$-sparse vector $\hat{\bf v}$ is obtained whose entries equal 1 for the $L$ largest correlation values, and zero otherwise.  
Finally, the decoded information word $\hat{\bf m}$ is obtained by inverse message-to-sparse vector mapping
	$\hat{\bf m} = {\cal S}^{-1}(\hat{\bf v})$.

\subsection{Long Codeword Decoding Complexity Issue}

The dictionary matrix
$\bf F$  in \eqref{eq:QOSSCenc} has $P N$ elements,  therefore the conventional decoder step~\eqref{eq:projection} has a complexity that scales as  $\mathcal{O}(P N)$. 
Compared to MDC ($L=1$)  in~\eqref{eq:Ninfo}, superposition ($L>1$) enables to transmit more information bits without increasing the dictionary size $N$. If $L$ is small, this can be done at marginal additional decoder complexity and performance degradation.  Reciprocally, for the same message size,  SSC can provide smaller decoding complexity by decreasing $N$ compared to ML decoding of MDC, and can thus enable feasible decoding in a larger range of codeword lengths $P$. 

Nevertheless, for a fixed $L$, to maintain a constant code rate $N_{\rm info}/P$ as $P$ increases, the dictionary size needs to grows exponentially with $P$, and so  ultimately decoding of long codewords becomes prohibitive also with SSC.  
For illustration, if considering a transmission on 64 PRBs in a slot of 14 OFDM symbols, the codeword length is of the order of  $P\simeq 168\times 64\simeq 10^4$ complex elements, leading  to typical  overall complexity order of $10^8 \sim 10^{12}$.

\section{Wideband Pilot-less ZC-QO-SSC}
\label{sec:wbQOSSC} 
A dictionary is said to be a QO set  if constructed as the union of multiple orthogonal subsets 
where the correlation between any  vectors in different orthogonal subset sets is bounded~\cite{KumarQOTIT00,StrohmerTIT06,bib:Pop2011ITW}.  
In~\cite{bib:Pop2011ITW}, we obtained a QO dictionary  
as the collection of different cyclic shift versions of ZC sequences and different root indices.  
In this section, we engineer multiple mechanisms based on ZC-QO-SSC to enable feasible decoding and good performance of long SSC codewords in  wideband  allocation.

\subsection{ZC-QO Dictionary}
A ZC sequence of prime length $P$ and root index $r$ is defined
as
\begin{equation}
	\label{eq:zcseq}
	z_r(k) = e^{-j \pi \frac{r}{P} k(k+1)}, k=0,\ldots,P-1
\end{equation}
where $j=\sqrt{-1}$ and $r$ is any nonzero positive integer less than $P$. 

Prime-length ZC sequences have the following well known properties: i) the cyclic autocorrelation of~\eqref{eq:zcseq} is zero for all non-zero lags;
ii) two ZC sequences of same prime length and different root indices have a normalized correlation bounded by $1/\sqrt{P}$. 
Hence, we construct $\bf F$ as
\begin{equation}
	\label{eq:projMtx}
	[{\bf F}]_{k,n} = z_{r(n)} (k+s(n) \; {\rm mod}\, P)
\end{equation}
for $k=0,\ldots,P-1$, $n=0,\ldots,N-1$, where $[{\bf F}]_{k,n}$ denotes the element of $\bf F$ at the $k$th  row and $n$th column. 
The root index and cyclic shift, $r(n)$ and $s(n)$ respectively, 
depend of the $n^{\rm th}$ column of ${\bf F}$, 
defined as follows:
\begin{equation}
	\label{eq:rootAndShift}
	\begin{split}
		r(n) & = 1 + \lfloor n/P\rfloor, \\
		s(n) & = n  \; {\rm mod}\, P.
	\end{split}
\end{equation}

With $\bf F$ defined as in~\eqref{eq:projMtx} and~\eqref{eq:rootAndShift},
the message-to-sparse vector mapping $\cal S$ can be equivalently seen as
mapping the information word $\bf m$ to $L$ pairs of integers $(r_l,s_l),
l=1,\ldots ,L$, where $r_l$ and $s_l$ are, respectively, root indices and
cyclic shifts of the columns corresponding to non-zero values of $\bf v$.
Thus, the elements of the sparse vector $\bf v$ are defined as 
\begin{equation}
v_n = \begin{cases}
1 & \text{if } \exists l \text{ such that } r(n)=r_l \text{ and } s(n)=s_l\\
	0 & {\rm otherwise}
\end{cases}.
\end{equation}

The elements $\{ c(k)\}_{p=0}^{P-1} $ of the codeword ${\bf c} $ 
in~\eqref{eq:QOSSCenc} can thus be re-written as 
\begin{equation}
	{c}(k) = \frac{1}{\sqrt{L}} \sum_{l=1}^{L}z_{r_l} (k+s_l \; {\rm mod}\, P), 
\end{equation}
i.e., the 
superposition of $L$ sequences, each one being determined by a root index
$r_l$ and cyclic shift $s_l$.

With ZC-QO, simple decoding complexity reduction can be carried out by noticing that all columns in the same orthogonal subset with same root index  are cyclic shifts of each other. Then, the correlations of a received codeword with an orthogonal subset can be computed all at once in the frequency domain.  Since with ZC-QO, the order of the dictionary size is $N \sim P^2$, this can bring down the complexity order from $P^3$ to $P^2 \log_2 P$. 
Nevertheless, the complexity is still too large to make decoding affordable on 
bandwidths larger than, e.g., 100 PRBs.

\subsection{ZC-QO-SSC With Embedded Data Root Indication}  
To significantly reduce  the decoding complexity,   
we propose a modified ZC-QO-SSC scheme with embedded data root index indication.

\subsubsection{Embedded Data Root Indication}

First, %
as previously the information word $\bf m$ is
mapped to $L$ pairs of root indices and cyclic shifts $(r_l,s_l)$, corresponding to $L$ ZC data sequences. 
However here in addition, a second set of $L$ indicator ZC sequences are selected, all from the same orthogonal subset with indicator root index $\bar{r}$. This root index $\bar{r}$ can be predefined and reserved among the root indices as typically the message-to-sparse vector mapping $\cal S$ leaves few root indices unused.  
Then, the cyclic shifts of this indicator ZC sequences are chosen to match the data root indices, i.e. $r_1,\ldots,r_L$.

The above $2L$ sequences  are
superposed, possibly with imbalanced power between  the $L$ data sequences and the $L$ indicator sequences. 
The SSC codeword  ${\bf c}=(c[0],\ldots,c[P-1])^T$ becomes
\begin{multline}
	\label{eq:wbCodeword} 
	c(k) = \frac{1}{\sqrt{L}}  \sum_{l=1}^L \left[\sqrt{\alpha} \	
		z_{r_l} (k+s_l \; {\rm mod}\, P) \right. \\ \left. + \sqrt{1-\alpha}\ z_{\bar{r}} (k+r_l \; {\rm mod}\, P) \right]
\end{multline}
\noindent where $\alpha$ is a power-splitting factor 
 between the data-carrying
sequences and indicator sequences. This parameter provides an additional degree of freedom to trade off indicator sequence detection and phase estimation accuracy against data sequence detection performance.

\subsubsection{Decoding}
Decoding of ZC-QO-SSC with embedded data root index indication is performed a follows. 
First, the decoder determines the data-carrying orthogonal subsets 
by identifying the cyclic shifts of the ZC sequence with indicator root index $\bar{r}$  returning the largest correlations. 
We select $L'\geq L$ cyclic shifts to avoid error propagation and to increase the chance of successful decoding.

For each set of $L$ candidate data root indices among the $L'$ selected above, we use the corresponding indicator sequences  $\{\mathbf{z}_l^{(I)}\}_{l=1}^L $ (with entries $z_l^{(I)}(k) =z_{\bar{r}} (k+\hat{r_l} \; {\rm mod}\, P) $)  as pilots to perform a global channel phase estimation and use this estimation to improve detection of the data sequences. Note that the indicator sequences are not exactly pilots as we do not know them precisely, but they are known to belong to a small orthogonal set which facilitates their detection.
We assume the channel approximately constant over a codeword such that if ignoring the noise, the received signal is $	{\bf y} \approx h {\bf c}$. 
We can thus have an estimate of the global channel as $	\hat{h} = \sum_{l=1}^L \left(\mathbf{z}_l^{(I)}\right)^H  {\bf y}$, up to a scaling factor. 
Even if the indication sequences would be perfectly estimated, this channel estimation would be corrupted by noise and interference from cross-correlation among ZC sequences with different roots. However the averaging over $L$ indication sequences mitigates these impairments. 
We use only the phase of global channel estimate $\angle 	\hat{h}$, to rotate the received signal as $\hat{\mathbf{y}} = {\bf y}e^{-j\angle 	\hat{h}} $, to perform coherent detection of the data sequences, i.e. sequences would be found as the ones maximizing the real part of the innner product as $\arg \max_{\mathbf{z}}  \Re \left \{ \mathbf{z}^H \hat{\mathbf{y}} \right\}$, instead of its magnitude as $\arg \max_{\mathbf{z} }  \lvert \mathbf{z}^H \hat{\mathbf{y}} \rvert$~\cite{bib:ChePitDirectSatellite}.

After this, the $L$ indicator sequences are canceled from the received signal, and for each candidate data root index the decoder identifies the cyclic shift that  produces the largest correlation. When a data sequence is identified, it is sequentially canceled in the received signal before detection of the next one. Once all data sequences are detected, the detector determines the remaining received signal energy. 

The process is repeated for all possible subsets of $L$ indicator sequences among the $L'$ detected  indicator sequences, and final detector output is selected according to the smallest remaining received signal energy.  

The decoding complexity is reduced as for determining a data sequence, a search is performed only in an orthogonal set corresponding the indicated root index, for a complexity of $P\log_2P$ compared to $P^2$. As only $L'$ data root indices are  considered in addition to the indicator root index, the decoding complexity is  $(L'+1) P \log_2 P$.

\subsection{Wideband  Multi-Codeword}
Any wireless channels, even line-of-sight (LoS) NTN ones, can experience frequent deep fading events.   
To improve robustness against deep fading, 
we transmit each SSC codeword
$R$ times -- that is, a first transmission followed by $(R-1)$ repetitions --
according to a repetition parameter $R$.

Retransmissions are done in different, non-contiguous time slots according to a time 
pattern aimed at maximizing the decoding success by exploiting time diversity.
For simplicity, we consider a pattern consisting of a set of uniformly spaced slots 
by a  time interval $T_R$ [slots] between consecutive transmissions.  
Thus, at the $t^{\rm th}$ slot we form a wideband SSC multi-codeword $	{\bf c}_W^{(t)}$ 
by concatenating the last encoded codeword ${\bf c}^{(t)}$ and repetitions
of previously generated codewords as 
$ {\bf c}_W^{(t)} = [ {\bf c}^{(t)}, {\bf c}^{(t-T_R)} , \cdots , {\bf c}^{(t-(R-1)T_R)}]^T $.   
The wideband multi-codeword $	{\bf c}_W^{(t)}$ is mapped to the REs
of the OFDM time-frequency grid according
to a frequency-first order as in 5G NR given that NTN channels are mostly frequency-flat, mapping codewords over the whole signal bandwidth.   
Time-first mapping could alternatively be considered for  frequency-selective channels, mapping then each codeword to a sub-band. 

At the receiver, repetitions of the same codeword are collected and combined. 
To improve combining of repetitions, we exploit the coarse channel estimation computed in the decoder of ZC-QO-SSC with embedded data root indication, discussed in the previous subsection. As such, the combiner performs  maximum ratio combining based on the average
channel phase and average received power of each transmission.

\subsection{Stop-Feedback}
In order to further improve performance we  
complement the wideband repetition mechanism 
with a stop-repetitions feedback message that informs the transmitter when a successful
decoding occurred.
Specifically, when the first transmission of a given codeword is in a non-faded slot,
most likely the decoder will be able to perform successful decoding based on the first
transmission alone. 
If decoding succeeds, the receiver sends back to the UE a stop-repetition feedback signal
 to indicate that the codeword has been successfully decoded and no
further repetitions are needed.
The UE, upon reception of that signal, stops transmission of further repetitions.  

Therefore, with stop feedback, each SSC codeword may be transmitted less than
$R$ times, thus making available the corresponding REs for possible reuse by other
codewords, either a new codeword carrying new data or additionnal repetitions for other past codewords. The former would increase throughput while the latter would reduce block error rate (BLER). .

The above stop-feedback message, although having obvious similarities with
conventional HARQ, has differences making it more suitable for satellite transmissions.
HARQ is based on transmissions of ACK/NACK feedback; NACK triggers transmission
of repetitions or incremental codeword redundancy.
However, the long round-trip time of satellite links causes HARQ stalling   
where awaiting for NACK results in long transmission times.  
Unlike conventional ACK/NACK, here the  repetitions is according
to a planned pattern, without awaiting reception of NACK,  making transmission faster. At the same time, by a stop-feedback signal, the
transmitter avoids sending unnecessary repetitions of a codeword that was successfully
decoded, improving overall the throughput.

\section{Performance Evaluations}
\label{sec:perfEval}

\paragraph{Simulation Assumptions} 
Performance evaluations were conducted  
using the same simulation framework as in~\cite{bib:ChePitDirectSatellite}. 
Transmit signal is	assumed to be operated at 2 GHz carrier frequency with a subcarrier spacing of 15 kHz. The channel follows the 3GPP NTN-TDL-C LoS model, with a delay spread of 3.5 ns and a UE velocity of 3 km/h.  
Residual frequency offset is neglected\footnote{The frequency offset can be largely compensated during initial access and further reduced by the receiver phase-tracking loop.  
For GNSS-assisted UEs, 3GPP TS 38.101-5 specifies a frequency accuracy of $\pm 0.1$ ppm. 
In~\cite{bib:ChePitDirectSatellite}, residual frequency offsets within this range were shown to have a negligible impact on demodulation performance at a carrier frequency of 2 GHz.}.   
The proposed ZC-QO-SSC scheme employs $L=2$ superpositions.

The BLER is computed as the ratio of number of decoded blocks with errors
over total number of transmitted blocks, the throughput $\Gamma$ follows as:
\begin{equation}\label{eq:Throughput}
\Gamma = (1-{\rm BLER}) N_{\rm info} \quad \text{[bits/s/Hz]}. 
\end{equation}

The SNR is computed in accordance to a Carrier-to-Noise Ratio (CNR) of -2.15 dB. This corresponds to a coverage-limited scenarios with a LEO-600 satellite system with parameter set 1 and elevation 30°, bandwidth 180 kHz (1 PRB),
a UE TX power of 23 dBm~\cite{bib:R1-2205856}. %
As the TX power remains constant regardless of the bandwidth, the SNR is obtained as
\begin{equation} \label{eq:SNR}
{\rm SNR} = -2.15 - 10 \log_{10} N_{\rm PRB} \quad \text{[dB]}
\end{equation} 
where $N_{\rm PRB}$ is the number of PRBs used for transmission.

\begin{figure}[!htbp]
	\centering 
	\vspace{0.0cm}
	\subfigure[BLER]{\includegraphics[width=.5\textwidth]{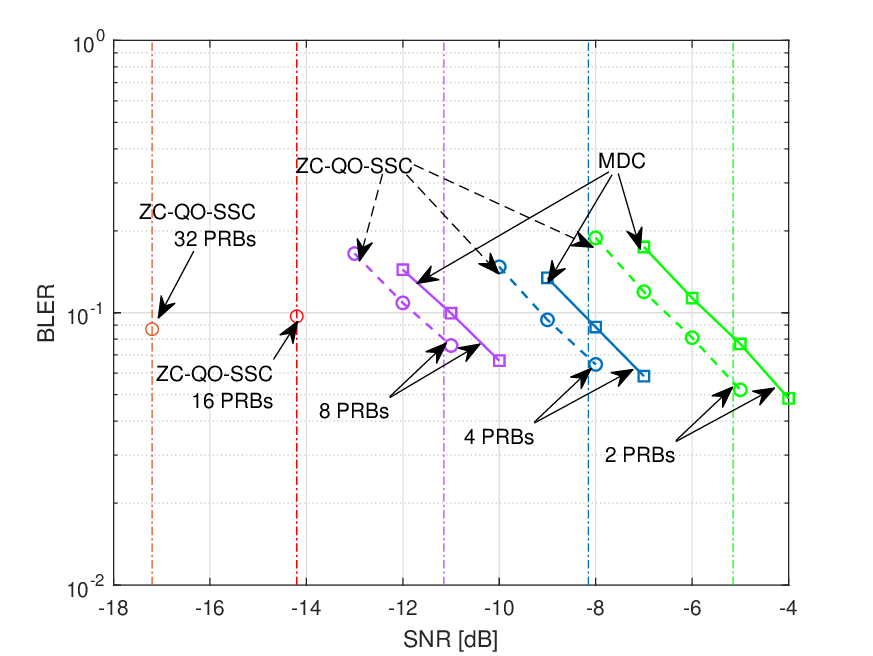}}  
\subfigure[Throughput]{\includegraphics[width=.5\textwidth]{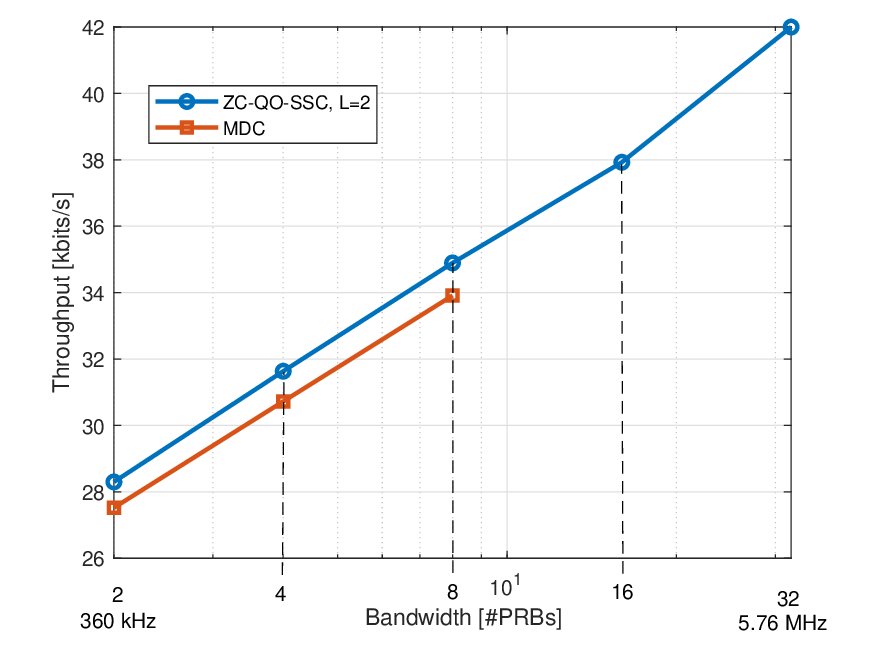}}      
\vspace{-0.1cm}
	\caption{Comparison of MDC and ZC-QO-SSC on NTN-TDL-C channel.}
	\vspace{-0.3cm}
	\label{fig:MDC_QOSSC_comp}
\end{figure}

\paragraph{MDC versus ZC-QO-SCC}
Fig.~\ref{fig:MDC_QOSSC_comp} compares the BLER between the pilot-less MDC approach in~\cite{bib:ChePitDirectSatellite} and the proposed pilot-less ZC-QO-SSC; with corresponding throughput~\eqref{eq:Throughput} at cell-edge SNRs~\eqref{eq:SNR} (indicated by dotted vertical lines on Fig.~\ref{fig:MDC_QOSSC_comp} (a)).  
The codeword length is $168 \times  N_{\rm PRB} $ with $ N_{\rm PRB} = 2, 4 , 8, 16, 32$, for which the information message lengths are respectively $K = 32, 36, 38, 42, 46$ [bits].
Since for MDC, ML decoding is prohibitively complex, as in~\cite{bib:ChePitDirectSatellite}, the  $K$ information bits are mapped to two MDC codewords allocated each to $N_{\rm PRB}/2$ PRBs.  

With ZC-QO-SSC, all the REs are used by a single codeword  carrying $K$ information bits by the superposition two quasi-orthogonal ZC sequences. 
 ZC-QO-SSC shows a 0.9-1 dB SNR gain over MDC, which is due to the fact that MDC needs to be segmented for feasible decoding.  
For bandwidths of 16 PRBs and 32 PRBs, MDC was not evaluated because of its complexity.

\begin{figure}[t]
	\centering 
	 \renewcommand{\figurename}{Tab.}
	 \renewcommand{\thefigure}{1}
	\vspace{0.4cm}
	\includegraphics[width=.49\textwidth]{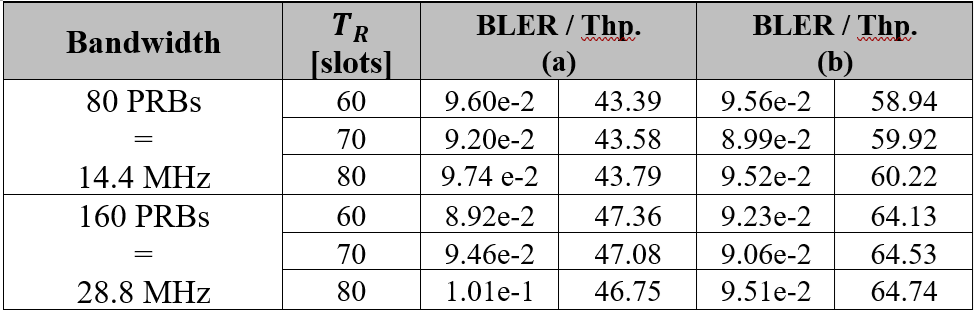}
\vspace{-0.2cm}
	\caption{BLER and throughput [kbits/s] of wideband ZC-QO-SSC with two repetitions. (a) no stop feedback; (b) Stop feedback with RE reuse.}
	\vspace{-0.0cm}
	\label{tab:EvalResultsTab}
\end{figure}

\paragraph{Wideband ZC-QO-SSC}
We now evaluate wideband transmission over 80 and 160 PRBs using the proposed ZC-QO-SSC scheme with embedded data root indication, repetition, and stop-feedback. A balanced power-splitting factor of $\alpha =1/2$ in~\eqref{eq:wbCodeword} and the  minimum repetition $R=2$ is adopted; optimizing these parameters is left for future work.   
Decoding employs $L'=7$, and different repetition periods $T_R$ are considered.  
Tab.~\ref{tab:EvalResultsTab} reports the results with and without stop-feedback. 
As in the previous evaluations, the maximum achievable throughput satisfying a BLER target of $10^{-1}$ at the cell-edge CNR is reported. According to~\eqref{eq:SNR}, this corresponds to SNRs of approximately $-21$ and $-24$ dB for $80$ and $160$ PRBs, respectively.  

Wideband ZC-QO-SSC achieves higher throughput as the allocated bandwidth increases.   
Unlike conventional binary coded modulation schemes, which typically become inefficient at extremely low code rates and very low SNRs required to exploit wider bandwidths under a fixed transmit power budget, the proposed ZC-QO-SSC effectively scales with bandwidth, achieving higher throughput despite the reduced operating SNR. 
We attribute this improvement to lower cross-correlation among superposed ZC sequences in longer SSC codewords,  together with better frequency diversity in wider bandwidths.  
Compared with the maximum throughput achieved in Fig.~\ref{fig:MDC_QOSSC_comp} using ZC-QO-SCC over  32 PRBs without embedded data root indication and repetition (i.e., $\alpha=1,\, R=1$),  the proposed enhancements increase the throughput by 11\% without stop-feedback and  by up to 54\% with stop-feedback.  

Overall, the achieved throughput ranges from 43 to 64 kbps. For comparison, the 3GPP Rel-18 study on NTN coverage enhancement reported support for low-data-rate services of 3 kbps in some scenarios, whereas 100 kbps could not be supported in any scenario~\cite{bib:R1-2208269}.  In~\cite{bib:ChePitDirectSatellite}, an NR-like reference scheme based on LDPC coding and QPSK achieved a maximum throughput of 18 kbps with one PRB and only 15.3 kbps with two PRBs under the same operating conditions. 
The MDC scheme proposed in~\cite{bib:ChePitDirectSatellite} increased the maximum throughput to 30.6 kbps over two PRBs.  
By enabling efficient operation over wider bandwidths, the proposed ZC-QO-SSC scheme further doubles the achievable throughput, reaching a maximum of 64 kbps, corresponding to a 3.5-fold throughput increase over the NR-like reference scheme in~\cite{bib:ChePitDirectSatellite}.

\section{Conclusions}
\label{sec:Conclusions}
This paper proposed a pilot-less coded modulation scheme for coverage-limited direct satellite uplink based on sparse superposition coding (SSC).  
Exploiting the Zadoff–Chu quasi-orthogonal  (ZC-QO) dictionary structure, its enables efficient decoding of long codewords. 
A  superposition encoding with root indicator ZC sequences was introduced to facilitate decoding while providing coarse channel estimates. 
In addition,   
a multi-codeword transmission with repetition and stop-feedback was developed, enabling improved decoding without conventional pilot signaling.  
These mechanisms allows a scalable SSC deployment, extending to wideband the pilot-less multi-dimensional constellation approach.  
Simulation results demonstrate that the proposed scheme significantly improves throughput by enabling wider bandwidth transmission compared to a multi-dimensional constellation scheme.  

\newpage
\section*{Acknowledgment}
The authors thank Dr. Erkai Chen of Huawei Technologies 
for developing  the MDC NTN simulator and for his help in integrating  
the proposed QO-ZC-SSC scheme.

\bibliographystyle{IEEEtran}
\bibliography{xbib}

\end{document}